# Tie-Line Characteristics based Partitioning for Distributed Optimization of Power Systems

A. Mohammadi, *Student Member, IEEE,* M. Mehrtash, *Student Member, IEEE*, A. Kargarian, *Member, IEEE*, and M. Barati, *Member, IEEE*

*Abstract—*The convergence performance of distributed optimization algorithms is of significant importance to solve optimal power flow (OPF) in a distributed fashion. In this paper, we aim to provide some insights on how to partition a power system to achieve a high convergence rate of distributed algorithms for the solution of an OPF problem. We analyzed several features of the power network to find a set of suitable partitions with the aim of convergence performance improvement. We model the grid as a graph and decompose it based on the edge betweenness graph clustering. This technique provides several partitions. To find an effective partitioning, we merge the partitions obtained by clustering technique and analyze them based on characteristics of tie-lines connecting neighboring partitions. The main goal is to find the best set of partitions with respect to the convergence speed. We deploy analytical target cascading (ATC) method to distributedly solve optimization subproblems. We test the proposed algorithm on the IEEE 118-bus system. The results show that the algorithm converges faster with a proper partitioning, whereas improper partitioning leads to a large number of iterations.

*Index Terms—* Edge Betweenness clustering, decentralized algorithm, Feature extraction, optimal power flow.

## I. Introduction

IMPLEMENTATION of distributed optimization algorithms is an alternative to the conventional centralized methods for power systems operation and planning[1]. This is incentivized by two main reasons, 1) information privacy in smart grids, and 2) distributing the computational burden on several processors. Moreover, distributed/decentralized algorithms potentially increase power systems reliability against failures of components or communication links.

Various distributed and decentralized optimization algorithms have been proposed in the literature to solve power system optimization problems. References[2]. provide a comprehensive literature review on the distributed/decentralized optimization algorithms and their applications on power systems. The main focus of this paper is on optimal power flow (OPF) that is a critical energy management function in a power system. Alternating direction method of multipliers (ADMM)[3, 4], auxiliary problem principle (APP)[4] [5], optimality condition decomposition (OCD)[6], consensus+innovation[7], and analytical target cascading (ATC)[8] are among the popular methods to solve OPF in a distributed/decentralized fashion. These methods coordinate a set of local OPF subproblems each of which is formulated for an area of the system. If the areas are known (based on the territory of a control entity), the coordination algorithms are applied to coordinate the local OPF problems. However, if the goal is to reduce the computational burden of the centralized OPF taking advantage of a distributed computing technique, the power system need to be decomposed into several small zones[9]. Then, a coordination strategy is applied to coordinate optimization subproblems of the zones.

The way that the system is decomposed into a set of zones has a significant impact on the convergence performance of the distributed optimization[10]. One approach is to partition the system to equal subsystems to benefit from the parallel computing. Several techniques, such as tableau[11], genetic algorithm[12], dynamic programming [13], and harmony search[14], have been presented to partition the system to a set of equal-sized subsystems. Although having equal-sized subproblems balances the computational cost of processors, it might increase the required iterations to achieve the convergence. This, consequently, increases the overall computational time.

Finding a proper partitioning that leads to a high convergence rate and accuracy is difficult. Indeed, not only the computational cost of each subproblem is important but also interdependencies of the subproblems and sensitivity of a zone to its neighboring zones play a critical role in the convergence rate and number required iterations. A proper partitioning depends on the system configuration, the number of buses in each zone, the number of tie-lines interconnecting the zones, the amount of power exchanged between the neighboring zones, the amount of load in each zone, etc.

In this paper, we analyze the impact of power system partitioning on the distributed OPF algorithm. We also provide some insights on how to find proper partitions to speed up the convergence rate and reduce the number of iterations of the distributed OPF algorithm. The grid is modeled as a graph (buses act as nodes and lines are edges) and implement the edge betweenness graph clustering to decompose the system into a set of zones. We partition the system in a way that each zone includes a subset of nodes and edges that are strongly connected. The 118-bus test system is partitioned in the different ways. We analyze several possible partitioning forms based on features of tie-lines interconnecting the zones and the convergence speed. Correlations between tie-lines' characteristics and the convergence speed are studied to provide some useful insights on power system decomposition.

## II. System Decomposition with Edge-Betweenness

To decompose the system into several subproblems, we use the optimal values from the centralized optimization. Then, the impact of different partitioning forms is investigated on the distributed algorithm.

### A. The Classical Centralized DCOPF

For the sake of explanation and simplicity, a DCOPF problem is considered.

This project was supported by the Louisiana Board of Regents under grant LEQSF (2016-19)-RD-A-10.
The authors are with the department of Electrical and Computer Engineering, Louisiana State University, Baton Rouge, LA 70803 USA (email: amoha39@lsu.edu, mmehrt3@lsu.edu, kargarian@lsu.edu, mbarati@lsu.edu.)

$$\min \sum_{i=1}^{NG} \underbrace{a_i \cdot p_i^2 + b_i \cdot p_i + c_i}_{f_i(p)} \quad (1)$$

$s.t.$

$$h(x) = 0 \leftrightarrow \begin{cases} p_i - d_i = \sum_{j \in \tau_i} \frac{\theta_i - \theta_j}{X_{ij}} \quad \forall i & (2) \\ \theta_{ref} = 0 & (3) \end{cases}$$

$$g(x) \leq 0 \leftrightarrow \begin{cases} P_{L_{ij}}^{min} \leq P_{L_{ij}} = \frac{\theta_i - \theta_j}{X_{ij}} \leq P_{L_{ij}}^{Max} & (4) \\ P_i^{min} \leq p_i \leq P_i^{Max} \quad \forall i & (5) \end{cases}$$

where $a_i$, $b_i$ and $c_i$ are the cost coefficients of generator unit and $p_i$ and $d_i$ denotes the amount of generation and load of bus $i$. $NG$ is total number of generator. $\theta$ denote a bus voltage angle. $h: \mathbb{R}^o \to \mathbb{R}^q$ and $g: \mathbb{R}^o \to \mathbb{R}^q$ are sets of equality and inequality constraints. $P_{L_{ij}}$ indicates the amount of power flow between bus $i$ and $j$. $\tau_i$ is the set of indices of buses that are connected to bus $i$.

### B. Interconnecting OPF Outputs to System Graph

To find the proper partitions, we run the centralized DCOPF once to obtain the system information. We consider the grid as a graph model in which buses are considered as nodes and lines as edges (we interchangeably use the terms "node" and "bus" throughout the paper; similarly for "line" and "edge"). Weights of the edges are needed to construct the graph. Three approaches can be selected. First, if two nodes are connected via a transmission line, the weight of the corresponding edge between these two nodes is one, else it is zero. Second, to assign the priority to each connection, the weight of each edge is defined based on the reactance of the corresponding transmission line as $1/X$. This is a good approach to segregate the strong and week connectivity between lines, while the impact of loads and generation are neglected. In the third approach, we define the weights according to line flows. A line with a large amount of flow, which may play a crucial role in the system, has more priority than a line with low flow. We select the third approach to define the weights of the edges.

$$w_{ij} = \begin{cases} P_{L_{ij}} \text{ if bus } i \text{ and } j \text{ connected} \\ 0 \end{cases} \quad (6)$$

Since the amount of line flows may vary in a wide range, we normalize the weights by dividing them by the minimum line flow. This gives a more proper weight for a line.

### C. Partitioning by Edge Betweenness

Many approaches can be deployed to partition the graph in various ways. Our goal in the decentralized optimization is to decompose the DCOPF problem into a set of subproblems each for a partition. It is desirable that the subproblems are weakly connected. In other words, a region (subproblem) should be connected to other regions with a minimum number of tie-lines. In addition, the weights of tie-lines need to be considered to account for their importance.

The edge betweenness algorithm is matched with our purpose. In this algorithm, inspires from vertex betweenness[15], the communities are detected based on the edges that are mostly between them. The centrality of a node and the impact of each node on the network are measured. To distinguish which edges are most between other nodes, the vertex betweenness is generalized. The edge betweenness for one edge is defined as the number of shortest paths between two nodes. If there is more than one minimal path between two vertices, an equal weight is assigned to each path while the total weight of all paths must be equal to one. Consider two communities, each of which is strongly connected (locally), whereas they have few inter-group edges, and all of the paths pass through these intergroup edges. In this case, we have a high edge betweenness connection, and most flows go along these inter-group edges. By removing these edges, the network is decomposed into two smaller networks. For instance, in Fig. 1, we determine the edge betweenness for a graph. The connected edge between nodes 4 and 5 has the maximum edge betweenness since a flow from nodes 1-4 need to pass edge 4-5 to reach nodes 5-8. This edge act as the bridge between the two subsets. To clarify more how this algorithm works the pseudocode for an unweighted graph is shown below:

| Algorithm of Edge-Betweenness |
|---|
| 1: **while** N partitions not obtained |
| 2:   Find all shortest paths between each pair of nodes |
| 3:   Divide shortest paths weights by $\chi$ that is number of shortest paths for each pair of nodes |
| 4:   Remove the edge with the highest betweenness |
| 5:   Recalculate the edge betweenness, go to step 2 |
| 6: **end while** |

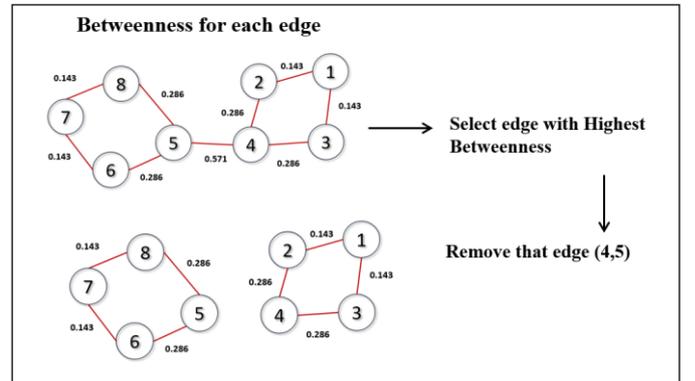

Fig. 1. Edge betweenness procedure.

### D. Construct a Set of Similar Clusters

Several partitions are obtained by implementing the edge betweenness graph clustering on the power network. The number of regions and coupling variables (i.e., the number of tie-lines) impact the distributed optimization. To have fair analysis and discussions, we merge the clusters obtained from the partitioning method in different ways to construct a set of new partitions in which the number of regions and tie-lines would be the same.

## III. ATC COORDINATION STRATEGY

We implement the analytical target cascading (ATC) method to solve the OPF problems of the partitions in a distributed manner[1]. ATC works based on the concept of augmented Lagrangian relaxation and coordinates the subproblems sequentially. Subproblems are placed at different hierarchical levels. A subproblem in an upper level ($l$) acts as a parent for the connected subproblems in the lower level $l + 1$, while children in the lower level are not linked [16]. A parent is linked to its children through a set of shared variables, which are voltage angles of the buses placed at the boundaries of the partitions in the considered DC OPF problem. ATC iteratively solves the OPF problems of the partitions (one problem at a time) and updates the shared variables to achieve a feasible and optimal solution from the perspective of the whole grid.

---

[1] Note that, in this paper, we aim to study the impact of grid partitioning. Without loss of generality, we deploy ATC to coordinate OPF subproblems. However, one can use other distributed optimization algorithms, such as APP and ADMM instead of ATC.

*A. ATC Formulation*

We briefly explain the ATC method (see [16] for more details). Assume a compact form of the centralized OPF as follows:

$$\min_{X} f(X) \quad (7)$$

$$s.t. \quad g(X) \leq 0, h(X) = 0 \quad (8)$$

where $X$ is the set of all variables (i.e., $p$ and $\delta$) of the system. $f: \mathbb{R}^o \to \mathbb{R}$ is the objective function (see (1)). $h: \mathbb{R}^o \to \mathbb{R}^q$ and $g: \mathbb{R}^o \to \mathbb{R}^q$ are the sets of equality (see (2) and (3)) and inequality (see (4) and (5)) constraints. The optimization can be rewritten as follows with respect to the local OPF subproblems of the partitions and their shared variables:

$$\min \sum_{\forall m} \sum_{\forall n} f_{mn}\left(x_{mn}, t_{(m+1)d_1}, \dots, t_{(m+1)d_{i_{mn}}}\right) \quad (9)$$

$$s.t. \quad g_{mn}\left(x_{mn}, t_{(m+1)d_1}, \dots, t_{(m+1)d_{i_{mn}}}\right) \leq 0 \quad (10)$$

$$h_{mn}\left(x_{mn}, t_{(m+1)d_1}, \dots, t_{(m+1)d_{i_{mn}}}\right) = 0 \quad (11)$$

$$\forall n \in E_m \quad m = \{1, \dots, M\}$$

where subscript $mn$ denotes subproblem $n$ in level $m$, $x_{mn}$ is the set of local variable of subproblem $n$ in level $m$. $t_{mn}$ is the set of shared variables (we call them target variables) defined in subproblem $n$ in level $m$. The target variables are determined by a parent and send down toward the corresponding children. $E_m$ is the set of subproblems in level $m$, and $d_{i_{mn}}$ is the number of subproblems in level $m$, and $M$ denotes the number of levels.

To separate the parents' and children's subproblems, a response copier $r$ is introduced for each targte variable $t$. The response variables are duplication of the targets. Targets and responses are shared variables between parents and children. The targets are controlled by parent, while the responses are handled with children. To enforce the decentralize algorithm to converge to the optimal (and feasible) point, the following set of consistency constraint must be satisfied:

$$C_{mn} = t_{mn} - r_{mn} = 0 \quad (12)$$

where $C_{mn}$ is the consistancy constriants of subproblem $n$ in level $m$. To relax the hard consistency constraints, the concept of augmented Lagrangian relaxation is deployed and a set of penalty functions are added to each subproblem's objective function. We can rewrite the objective function (9) as:

$$\min \sum_{\forall m} \sum_{\forall n} f_{mn}(x_{mn}, t_{(m+1)d_1}, \dots, t_{(m+1)d_{i_{mn}}})$$

$$+ \pi_{mn}\left(c_{(m+1)d_1}, \dots, c_{(m+1)d_{i_{mn}}}\right) \quad (13)$$

subject to (10) and (11). $\pi_{mn}$ denotes the penalty term added to subproblem $n$ in level $m$. Several options exist to model the penalty terms, such as exponential and quadratic functions. With the above procedure, the partitions are placed in different levels and their OPF subproblems are separated.

*B. AL-AD Coordination Strategy*

Several algorithms can be deployed to coordinate the OPF subproblems. In this paper, we select a second-order penalty term and follow the alternating direction method of multipliers (AL-AD) coordination strategy[3]:

$$\pi_{AL-AD}(c) = \lambda^T(t - r) + \|\omega \circ (t - r)\|_2^2 \quad (14)$$

where $\lambda$ and $\omega$ are penalty multipliers and "∘" denotes the Hadamard product. In the DC OPF, the target and response variables are the voltage angles of terminals of tie-lines that connect the partitions. For the sake of simplicity and explanation, we consider two partitions, and put partition 1 in level 1 and partition 2 in level 2. In each iteration $k$ of AL-AD (it is an iterative procedure), OPF subproblem 1 in level 1 (i.e., the parent partition) is:

$$\min_{(x_{11}, \theta_{22})} f_{11}(x_{11}, \theta_{22}) + \lambda^T(\theta_{22} - \tilde{\theta}_{22}^{k-1})$$

$$+ \|\omega \circ (\theta_{22} - \tilde{\theta}_{22}^{k-1})\|_2^2 \quad (15)$$

where $\theta$ denotes the target variables of the parent, and $\tilde{\theta}$ is the response variables of the child. And the OPF subproblem 2 in level 2 (i.e., the child partition) is:

$$\min_{(x_{11}, \tilde{\theta}_{22})} f_{22}(x_{11}, \tilde{\theta}_{22}) + \lambda^T(\theta_{22}^{k-1} - \tilde{\theta}_{22})$$

$$+ \|\omega \circ (\theta_{22}^{k-1} - \tilde{\theta}_{22})\|_2^2 \quad (16)$$

Note that $\theta$ is constant in (16), wherase $\tilde{\theta}$ is constant in (15). The AL-AD 's steps are as follows:

Step1: Initialize local variables $x$ of each subproblem, target values $t$, response $r$, penalty multipliers $\lambda$ and $\omega$, and parameter $\beta$, and set the iteration index $k = 1$.

Step2: Solve OPF subproblems in level $l$, and update the target values. Solve subproblems in level $l + 1$, update response (for levels$< l + 1$) and target (for levels$> l + 1$) values. Do that until all levels are solved.

Step3: If $\max(\|\theta^k - \tilde{\theta}^k\|) \leq \epsilon$, where $\epsilon$ is a stopping threshold, the convergence is achieved. Otherwise, $k \leftarrow k + 1$ and update the penalty multipliers as follows:

$$\lambda^k = \lambda^{k-1} + \omega \circ (\theta^{k-1} - \theta^{k-1}) \quad (17)$$

$$\omega^k = \beta \omega^{k-1} \quad (18)$$

and then go to Step 2.

We refer to[16] more details on the ATC method and the AL-AD coordination strategy.

## IV. NUMERICAL RESULTS AND DISCUSSIONS

The edge betweenness partitioning method is implemented on the IEEE 118-bus system, and the simulation results are discussed. We explain how the test system is decomposed into different clusters. We extract features of each cluster to understand which cluster is more efficient. The AL-AD coordination strategy is solved to analyze the accuracy of the clustering framework. To have a fair analysis and comparison, the convergence criterion and initial points are the same in all cases. All simulations are carried out using R and MATLAB, on a personal 2.7 GHz computer with 16 GB RAM.

### A. Decomposing 118-Bus System

To partition the test system, the centralized DC OPF problem is solved to determine the line flows. Based on our experience, we cluster the system to eight zones. Figure 2 (a) shows the nodes in each zone. For the sake of analysis and discussions and to have a fair comparison between the partitions, we combine the partitions in different ways to create two zones that are connected through three tie-lines. We find four cases and, in Fig. 2 (b), show how clusters are connected. The weight of each edge denotes the number of tie-lines between two clusters. Each of the dash-line separates the graph into two parts (e.g., {3,5} are in the first group, and {1,2,4,6,7,8} are in the second group). The four cases are given in Table I.

### B. Tie-line Features

We find the edges placed between the clusters and define them as bridges. By removing the bridges, the clusters are completely isolated. Since each bridge is placed in a different part of the power grid and has different reactance, they have different features. We define two indies to characterize the features of each bridge. The first index is the bus voltage angles variations. This index indicates that if we change the voltage angles of two sides of a bridge, how other angles change. This measures the importance of each bridge. If the impact of this variation is high, it means that this bridge it is not a proper tie-lines, as a small change in this tie-line has a high impact on the rest of the system, and the distributed algorithm put more efforts to find the optimal point for both sides of this tie-line. To calculate this index, we solve the problem in a centralized manner. Knowing the optimal angles, we increase/decrease the two sides of the bridge by 0.1%. We then solve the problem again to observe variations of the voltage angles in the rest of the system. The angle variation is formulated as:

$$rel_\theta = \frac{|\sum_i \theta_i^{mod} - \sum_i \theta_i^{opt}|}{|\sum_i \theta_i^{opt}|} \quad (19)$$

where $\theta_i^{mod}$ denotes the voltage angle of bus $i$ when the angles of the bridge terminals vary. $\theta_i^{opt}$ is the optimal value for the angle of bus $i$.

The second index measures the impact of a bridge on the cost function. The procedure of calculation of the cost variation is the same as of $rel_\theta$. This index measures how much the cost function vary.

$$rel_f = \frac{|\sum_i f_i^{mod} - \sum_i f_i^{opt}|}{|\sum_i f_i^{opt}|} \quad (20)$$

where $f_i^{mod}$ and $f_i^{opt}$ are respectively the cost of generation unit $i$ after the angle modification and before that (i.e., the optimal angles). Note that in the four partitioning shown in Table I, the two zones are connected via four tie-lines. We add the defined indices of each tile-line to find two equivalent indices.
To analyze the impact of power demand, we examine three tests with the normal load, 75% of the normal load (low load), and 125% of the normal load (high load). To provide a better sense, we normalize the result by dividing them by their average value. The results are shown in Tables II.

### C. Impact of Partitioning on Distributed Algorithm

AL-AD is applied to find DCOPF solution of the four possible partitioning cases under different loading conditions. To have a fair condition for all cases, the initial value for targets and responses are set to zero. The stopping criterion is $\epsilon = 5 \times 10^{-4}$, and $k \leq 100$. The initial values for multipliers are $\lambda = 500$, $w = 500$. All cases converge after 100 iterations.

We define two indices, to evaluate our results. The first index is the relative distance of the total cost determined by centralized OPF and the decentralized one.

$$rel_{cost} = \frac{|f_{cent} - f_{dec}|}{f_{cent}} \quad (21)$$

where $f_{cent}$ and $f_{dec}$ are the optimal cost function obtained by the centralized and decentralized OPF. The second index, shows the average of $rel_{cost}$ over the course of iterations. In several iterations, $rel_{cost}$ is precise but the performance of AL-AD in its previous iterations is not good. Hence, we formulate the $rel_{avg_{cost}}$ as

$$rel_{avg_{cost}} = \frac{\sum_{k=1}^{K} \frac{|f_{cent} - f_{dec}^k|}{f_{cent}}}{K} \quad (22)$$

where $f_{dec}^k$ is the value of the objective function of the distributed algorithm in iteration $k$. Table III shows the results of the four possible partitioning forms. The results demonstrate that a case with small $rel_\theta$ and $rel_f$ would lead to a better result. However, the impact of $rel_\theta$ is higher than $rel_f$. We use a fuzzy logic to assign one value instead of using both $rel_\theta$ and $rel_f$ [17]. If we consider Case 2 in the normal load, the $rel_\theta$ and $rel_f$ indices are less in comparison with other cases. The results prove that Case 2 has proper values in comparison with the three other cases. On the other hand, $rel_\theta$ for Case 1 in the normal load is the worst. Thus, we expect that this case would be the worst case.

To investigate the effect of $rel_f$ and $rel_\theta$ the our results, we select various amount of load as {0.6, 0.7, 0.8, 0.9, 1, 1.1, 1.2}, and test Case 2. The features of each partion are extracted and normalized (divide by the mean value). The results are ploted in Fig. 3 (a) based on $rel_f$ and $rel_\theta$, and the amount of $rel_{avg_{cost}}$ is depicted at each point. The result shows that the point with an small value of $rel_f$ ad $rel_\theta$ has the least error, while the worst case has $rel_f = 1.06$ and $rel_\theta = 3.1$. In Fig. 4 (b), $rel_{cost}$ for the worst and best cases are drawn (logaritmic) over the course of iterations.

### V. CONCLUSION

In this paper, we studied the impact of power grid partitioning on the distributed OPF. We deployed the edge betweenness partitioning approach to decompose the system. Several possible partitioning forms could be selected. The main concern is "which partitioning form is suitable for the decentralized OPF algorithm?" We introduced several indices obtained from features of the tie-lines connecting the partitions to analyze the results of distributed OPF of each partitioning case. The simulation results on the 118-bus system showed that each partitioning case provides different convergence performance. A partitioning case with the least variation indices is potentially the best case from the perspective of the decentralized algorithm, i.e., AL-AD shows a good convergence performance for such a case.

TABLE I. SUBSETS OF EACH CASE

| Case ID | subsets |
|---|---|
| 1 | {3 5}, {1 2 4 6 7 8} |
| 2 | {6}, {1 2 3 4 5 7 8} |
| 3 | {4 6 8}, {1 2 3 5 7} |
| 4 | {2 4 6 8}, {1 3 5 7} |

TABLE II. FEATURES OF EACH CLUSTER

| Case No | Normal load | | Low load | | High load | |
|---|---|---|---|---|---|---|
| | $rel_{cost}$ $\times 10^{-3}$ | $rel_{avg_{cost}}$ $\times 10^{-3}$ | $rel_{cost}$ $\times 10^{-3}$ | $rel_{avg_{cost}}$ $\times 10^{-3}$ | $rel_{cost}$ $\times 10^{-3}$ | $rel_{avg_{cost}}$ $\times 10^{-3}$ |
| 1 | 0.9052 | 2.0833 | 0.7854 | 0.3779 | 1.2733 | 2.7536 |
| 2 | **0.4491** | **0.1164** | 2.9915 | 1.0350 | 1.1451 | 0.2144 |
| 3 | **0.3793** | 0.3591 | 0.2108 | 2.3676 | **0.0620** | **0.0581** |
| 4 | 2.2664 | 1.4411 | **0.0121** | **0.2192** | 1.5193 | 0.9737 |

TABLE III. ERROR OF EACH CLUSTER

| Case No | Normal load | | Low load | | High load | |
|---|---|---|---|---|---|---|
| | $rel_{cost}$ $\times 10^{-3}$ | $rel_{avg_{cost}}$ $\times 10^{-3}$ | $rel_{cost}$ $\times 10^{-3}$ | $rel_{avg_{cost}}$ $\times 10^{-3}$ | $rel_{cost}$ $\times 10^{-3}$ | $rel_{avg_{cost}}$ $\times 10^{-3}$ |
| 1 | 3.2 | 7.1 | 0.09 | 2.8 | 5.4 | 8 |
| 2 | **0.07** | **1.5** | 1.21 | 3.0 | 1.35 | 3.2 |
| 3 | **0.01** | 3.2 | 0.01 | 3.4 | **0.004** | **3.1** |
| 4 | 0.02 | 7.5 | **0.007** | 7 | 0.22 | 8.3 |

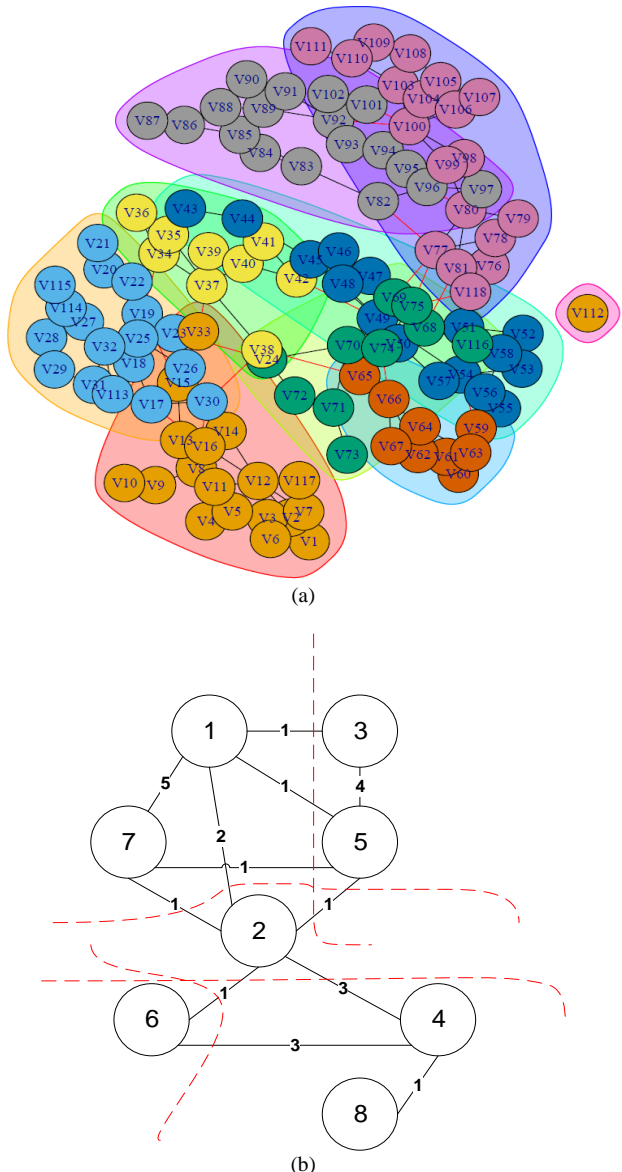

Fig. 2. a) nodes clustering and b) merge clusters.

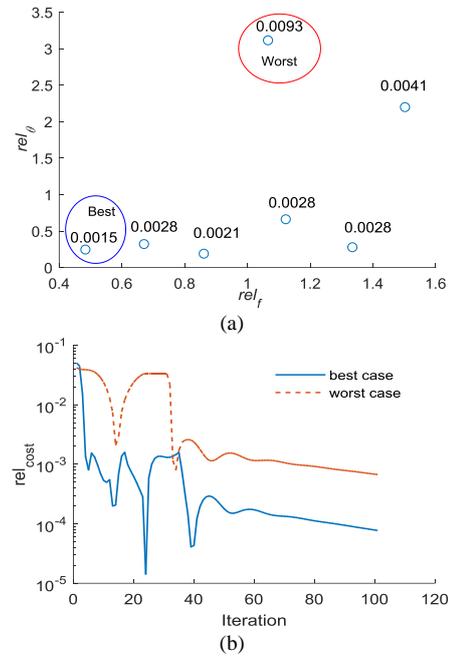

Fig. 3. a) Impact of $rel_f$ and $rel_\theta$ on decentralized algorithm b) $rel_{cost}$ over the course of iterations.